\documentclass{PoS}
\usepackage{lineno}
\usepackage{amsmath}
\usepackage{graphicx}

\title{Sensitivity of HAWC to Primordial Black Hole Bursts}

\ShortTitle{Sensitivity of HAWC to PBH Bursts}

\author{T. N. Ukwatta$^a$, J. T. Linnemann$^b$, D. Stump$^b$, \speaker{J. H. MacGibbon}$^c$, S. S. Marinelli$^b$, T. Yapici$^b$, and K. Tollefson$^b$ for the HAWC Collaboration$^d$ \\
        \llap{$^a$}Space and Remote Sensing (ISR-2), Los Alamos National Laboratory, Los Alamos, NM 87545, USA.\\
        \llap{$^b$}Department of Physics and Astronomy, Michigan State University, East Lansing, MI 48824, USA.\\
        \llap{$^c$}Department of Physics, University of North Florida, Jacksonville, FL 32224, USA.\\
        \llap{$^d$}For a complete author list, see \href{http://www.hawc-observatory.org/collaboration/icrc2015.php}{www.hawc-observatory.org/collaboration/icrc2015.php}\\
        Email: \email{tilan@lanl.gov}}

\abstract{Primordial Black Holes (PBHs) are black holes that may have been created in the early Universe and could be as large as supermassive black holes or as small as the Planck scale. It is believed that a black hole has a temperature inversely proportional to its mass and will thermally emit all species of fundamental particles. PBHs with initial masses of $\sim5.0 \times 10^{14}$ g should be expiring today with bursts of high-energy gamma radiation in the GeV/TeV energy range. The High Altitude Water Cherenkov (HAWC) observatory is sensitive to the high end of the PBH gamma-ray burst spectrum. Due to its large field of view, duty cycle above 90\% and sensitivity up to 100 TeV, the HAWC observatory is well suited to perform a search for PBH bursts. We report that if the PBH explodes within 0.25 light years from Earth and within 26 degrees of zenith, HAWC will have a 95\% probability of detecting the PBH burst at the 5 sigma level. Conversely, a null detection from a 2 year or longer HAWC search will set PBH upper limits which are significantly better than the upper limits set by any previous PBH search.}

\FullConference{The 34th International Cosmic Ray Conference,\\
		30 July- 6 August, 2015\\
		The Hague, The Netherlands}

\begin{document}

\section{Introduction}\label{intro}

In the current Universe, there are no known processes that can create black holes with masses less than a few solar masses. However, early in the Universe, conditions may have been extreme enough to create black holes with masses ranging from the Planck mass to super-massive black holes~\cite{Carr2010}. These black holes are called Primordial Black Holes (PBHs). PBH production in the early Universe can have observable consequences, spanning from the largest scales, for example influencing the development of large-scale structure in the Universe, to the smallest scales, for example enhancing local dark matter clustering. In the present Universe, PBHs in certain mass ranges may constitute a fraction of the dark matter~\cite{Carr2010}.

It is expected that every black hole thermally radiates (`evaporates') with a temperature inversely proportional to its mass~\cite{BH_Hawking}. The emitted radiation consists of all fundamental particle species with masses less than about the black hole temperature~\cite{MacGibbon1990}. PBHs with an initial mass of $\sim \, 5.0 \times 10^{14}$ g should be expiring now with bursts of high-energy particles, producing gamma radiation in the GeV -- TeV energy range~\cite{MacGibbon2008}. Positive detection of a PBH burst event would provide important information about the processes that shaped the
early Universe. In addition, confirmed detection will give access to particle physics at energies higher than currently achievable by terrestrial accelerators.
Even the non-detection of PBH burst events in dedicated searches would give important information about fundamental physics and the early Universe.

Numerous observatories have searched for PBH burst events using direct and indirect methods. These methods are sensitive to the PBH distribution
at various distance scales.
Observatories that observe photons or antiprotons at $\sim$100 MeV can probe the cosmologically-averaged or Galactic-averaged PBH distribution whereas TeV observatories directly probe PBH bursts on parsec scales.
Due to the possibility that PBHs may be clustered at various scales, all these searches provide important information.
We also note that the TeV direct search limits apply to any local bursting black holes, regardless of their formation mechanism or formation epoch.
Table~\ref{limit_table} gives a summary of various search methods, the distance scales they probe and their current best limits.

\begin{table}[h]
\begin{center}
\begin{tabular}{|l|c|c|}
\hline Distance Scale & Limit & Method \\ \hline
Cosmological Scale & $< \, 10^{-6}$ ${\rm pc^{-3} yr^{-1}}$ & (1) \\
Galactic Scale & $<$ 0.42 ${\rm pc^{-3} yr^{-1}}$ & (2) \\
Kiloparsec Scale& $<$ $1.2 \times 10^{-3}$ ${\rm pc^{-3} yr^{-1}}$ & (3) \\
Parsec Scale & $<$ $1.4 \times 10^{4}$ ${\rm pc^{-3} yr^{-1}}$ & (4) \\ \hline
\end{tabular}
\caption{PBH burst limits on various distance scales: (1) from the 100 MeV extragalactic gamma-ray background assuming no PBH clustering~\cite{Carr2010, PageHawking1976}, (2)
from the Galactic 100 MeV anisotropy measurement~\cite{Wright1996}, (3) from the Galactic antiproton flux~\cite{Abe2012} and (4) from Very High Energy direct burst searches~\cite{Glicenstein2013}.}
\label{limit_table}
\end{center}
\end{table}

In this paper, we present the sensitivity of the High Altitude Water Cherenkov (HAWC) observatory to PBH bursts, which
tries to improve the PBH limit at the parsec scale.

\section{HAWC Observatory}

The HAWC observatory is a very high energy (VHE) observatory located at Sierra Negra, Mexico at an altitude of 4,100~m above sea level. HAWC consists of 300 water tanks, each 7.3 m in diameter and 4.5 m deep. Each tank houses three 8-inch photo multiplier tubes (PMTs) and one 10-inch PMT. The PMTs detect Cherenkov light from secondary particles created in extensive air showers induced by VHE gamma rays of energies between $\sim$50 GeV and 100 TeV.
The main data acquisition system measures the arrival direction and energy of the VHE gamma ray by timing the arrival of the secondary particles on the ground and measuring the amplitude of the PMT signals.
The direction of the primary particle may be resolvable with an error of between 0.1 and 2.0 degrees depending on its energy and location in the sky. HAWC has a large instantaneous field-of-view of $\sim$ 2 sr and will have a high duty cycle of greater than 95\%. Thus, HAWC will be able to observe high-energy emission from gamma-ray transients. For more information about the HAWC observatory please see references~\cite{Pretz2015, Smith2015} and references therein.

\section{Methodology}\label{Methodology}

\subsection{Primordial Black Hole Burst Spectrum}

As the PBH radiates, it continually loses mass and its temperature increases to very high energies. The manner in which the PBH expires depends on physics at very high energies. In the Standard Evaporation Model (SEM)~\cite{MacGibbon1990, MacGibbon1991} (based on the Standard Model of particle physics), a PBH should directly radiate those fundamental particles whose Compton wavelengths are comparable with the size of the black hole. When the black hole temperature exceeds the Quantum Chromodynamics (QCD) confinement scale ($\sim$250--300 MeV), quarks and gluons should be radiated~\cite{MacGibbon1990, PageHawking1976}. The quarks and gluons will then fragment and hadronize as they stream away from the black hole, analogous to the jets seen in terrestrial accelerators~\cite{MacGibbon1990, MacGibbon2008}. On astrophysical timescales, the final jet products will be photons, neutrinos, electrons, positrons, protons and anti-protons. In this work, we assume the SEM as our emission and particle physics model. Please refer to reference~\cite{Ukwatta2015b} for more information about this model, and for the calculation of PBH burst light curves in the HAWC energy range and their energy dependence.

\begin{figure}
\centering
\includegraphics[width=0.7\textwidth]{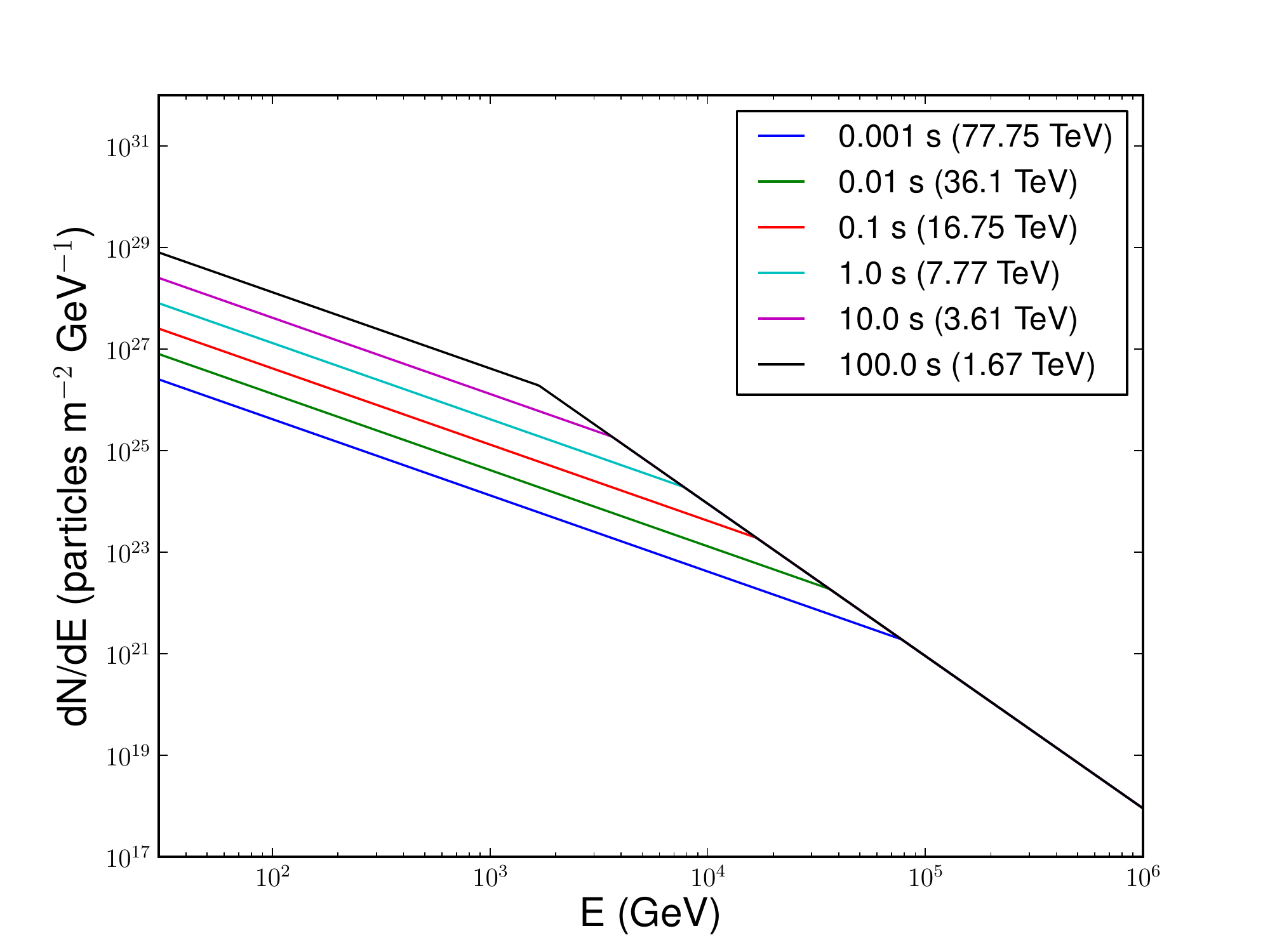}
\caption{Time-integrated gamma-ray spectrum over various PBH remaining lifetimes using the parametrization of Equation \protect\ref{photonEq}. The black hole temperature at the start of observation is shown in parentheses.}\label{pbh_spectrum}
\end{figure}

In the SEM model, the black hole temperature ($T$) can be expressed in terms of the remaining evaporation lifetime ($\tau$) of the black hole (that is, the time left until the black hole stops evaporating) as follows~\cite{MacGibbon1991,Petkov2008}:

\begin{equation} \label{tempEq}
T \simeq \bigg[4.8 \times 10^{11} \, \bigg(\frac{\rm{1 sec}}{\tau}\bigg) \bigg]^{1/3}\,\,\,\rm GeV
\end{equation}
\\
for temperatures well below its ultimate or Planck temperature. The emission rate increases as the black hole shrinks~\cite{PageHawking1976}. For black holes with temperatures greater than several GeV at the start of the observation, the time integrated photon flux can be parameterized as~\cite{Petkov2008}

\begin{equation} \label{photonEq}
\frac{dN}{dE} \approx 9 \times 10^{35} {\,\, \rm particles \,\, m^{-2} GeV^{-1}}
\begin{cases}
\big(\frac{1 GeV}{T}\big)^{3/2}\big(\frac{1 GeV}{E}\big)^{3/2},\,\,\,\,E<T \\
\big(\frac{1 GeV}{E}\big)^{3},\,\,\,\,E\ge T
\end{cases}
\end{equation}
\\
for gamma-ray energies $E \gtrsim$ 10 GeV. This parametrization includes both the directly radiated photons and those produced by the decay of other directly radiated species. Figure~\ref{pbh_spectrum} shows the total gamma-ray spectrum for various PBH remaining lifetimes ranging from 0.001 s to 100 s. The parametrization of the instantaneous PBH spectra is presented in reference~\cite{Ukwatta2015b}.

\subsection{PBH Upper Limit Estimation}

In the case of a null detection by HAWC, we can derive an upper limit on the local PBH burst rate density. The parametrization given in equation~\ref{photonEq} can be used to calculate the expected number of photons detectable by an observatory on the Earth's surface. For a PBH burst of duration $\tau$ at a non-cosmological distance $r$ and zenith angle $\theta$, the number of expected photons is

\begin{equation} \label{countsEq}
\mu(r, \theta, \tau) = \frac{(1-f)}{4 \pi r^2} \int_{E_1}^{E_2} \,\frac{dN(\tau)}{dE}\, A(E,
\theta)\,dE
\end{equation}
\\
where $dN/dE$ is the PBH gamma-ray spectrum integrated from remaining lifetime $\tau$ to 0. The energies $E_1$ and $E_2$ correspond to the energy range of the detector, $A(E,\theta)$ is the effective area of the detector as a function of photon energy and zenith angle, and $f$ is the dead time fraction of the detector.

The minimum number of counts needed for a detection for different burst durations $\tau$, $\mu_{\circ}(\theta_i, \tau)$, can be estimated by finding the number of counts required over the background for a $5 \sigma$ significance. We define a 5$\sigma$ detection after correction for $N_t$ trials to be the number of counts $n$ which would have a Poisson probability $P$ corresponding to a p-value $p_c$ given by

\begin{equation} \label{stat1}
p_c = p_o /N_t = P(\geq n|n_{\rm bk})
\end{equation}
\\
where $p_0 = 2.3 \times 10^{-7}$ (the p-value corresponding to 5$\sigma$), $n_{\rm bk}$ is the number of background counts expected over the burst duration $\tau$, and $P(\geq n|n_{\rm bk})$ denotes the Poisson probability of getting $n$ or more counts when the Poisson mean is $n_{\rm bk}$. We take the value of $\mu_{\circ}(\theta_i, \tau)$ to be the amount of expected signal which would satisfy this criterion 50\% of the time. To find $\mu_{\circ}(\theta_i, \tau)$, we estimate the value at which the Poisson probability of finding at least $n$ counts is 50\% by using the relation

\begin{equation} \label{stat2}
P(\geq n | (n_{\rm bk} + \mu_{\circ}(\theta_i, \tau))) = 0.5.
\end{equation}
\\
The maximum distance from which a PBH burst could be detected by HAWC can be calculated by
equating the values of $\mu_{\circ}(\theta_i, \tau)$ to $\mu(r, \theta_i, \tau)$ and solving for $r$,

\begin{equation} \label{distanceEq}
r_{\rm max}(\theta_i, \tau) = \sqrt{ \frac{(1-f)}{4 \pi \mu_{\circ}(\theta_i, \tau)} \int_{E_1}^{E_2} \,\frac{dN(\tau)}{dE}\, A(E,\theta_i)\,dE}.
\end{equation}
\\
The detectable volume from which a PBH could be detected is then

\begin{equation} \label{volueEq1}
V(\tau) = \sum_{i} V(\theta_{ i}, \tau) = \frac{4}{3} \pi \sum_{i} r_{\rm max}^3(\theta_{ i}, \tau) \frac{{\rm FOV_{\rm eff}}(\theta_{i})}{4\pi}
\end{equation}
\\
where $\rm FOV_{\rm eff}(\theta_{i})$ denotes the effective field-of-view of the detector for a given zenith band $i$,

\begin{equation} \label{fov}
{\rm FOV_{\rm eff}}(\theta_{i}) = 2 \pi (\cos \theta_{i,\,\rm min}-\cos \theta_{i,\,\rm max})\ {\rm sr}
\end{equation}
\\
and $\theta_{i,\, \rm min}$ and $\theta_{i,\, \rm max}$ are the minimum and maximum zenith angles in band $i$.

If we assume that the PBHs are uniformly distributed in the solar neighborhood, the
X\% confidence level upper limit ($UL_{X}$) on the local rate density of
PBHs bursts (that is, the number of bursts occurring locally per unit volume per unit time) can be estimated as

\begin{equation}\label{ulX}
UL_{X} = \frac{m}{V S},
\end{equation}
if, at the X\% confidence level, the detector observes zero bursts.
Here $V$ is the effective detectable volume from which a PBH can be detected, $S$ is the total search duration
and $m$ is the upper limit on the expected number of
PBH bursts given that at the X\% confidence level zero bursts are observed at Earth.
For Poisson fluctuations, $m = \ln (1/(1-X))$. If $X=99\%$, the upper limit on the PBH burst rate density will be

\begin{equation}\label{ul99}
UL_{99} = \frac{4.6}{V S}.
\end{equation}

\section{Results and Discussion}\label{Results}

\begin{figure}[htp]
\centering
\includegraphics[width=0.7\textwidth]{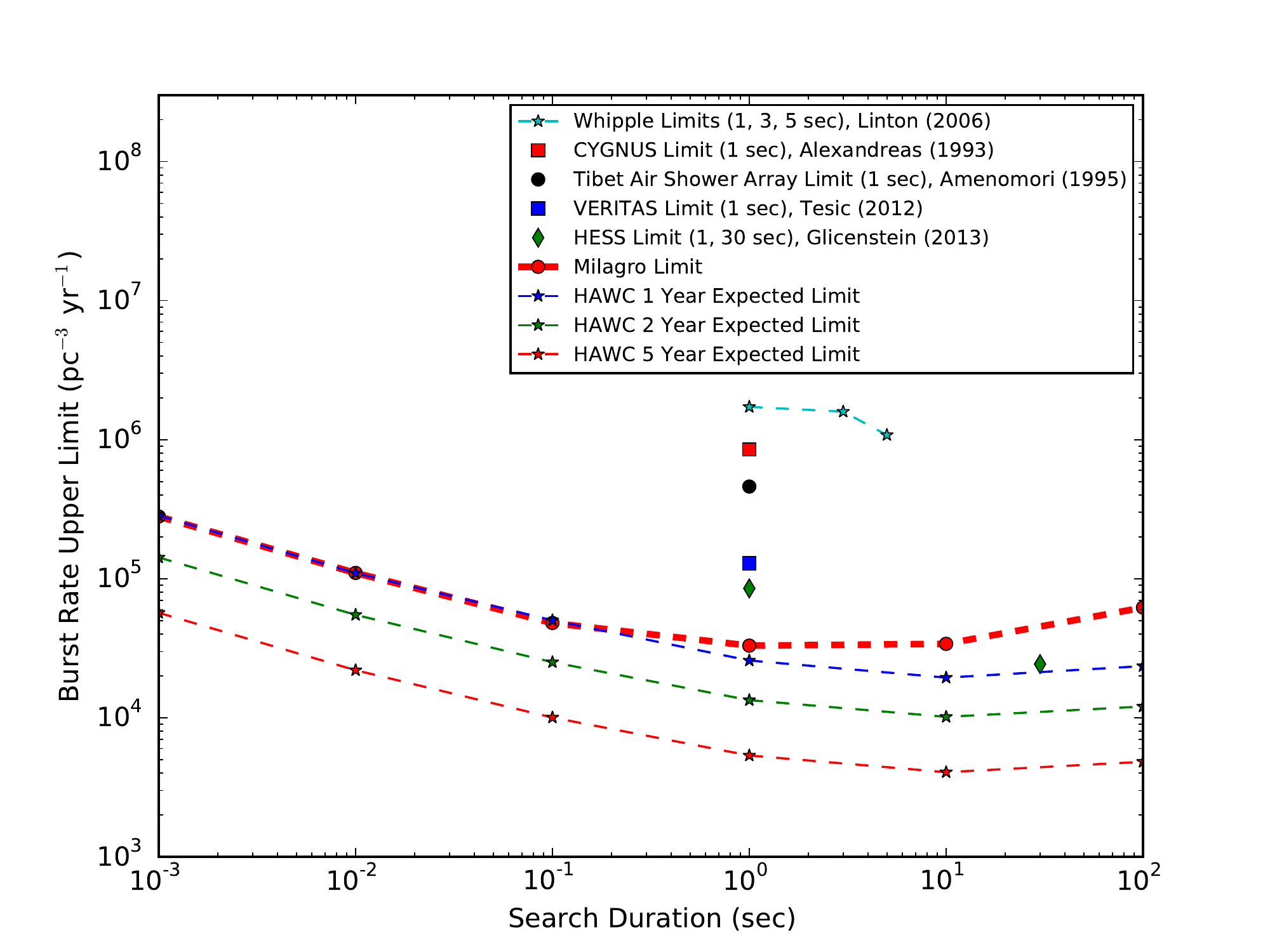}
\caption{PBH Burst Rate Density Upper Limits projected for HAWC~\cite{Ukwatta2015}, compared with limits from previous direct search experiments.}\label{pbh_limits}
\end{figure}

The methodology presented in Section~\ref{Methodology} can be used to find the projected PBH burst rate density upper limits for HAWC in the case of null detection in a 5 year search. We employ a Monte Carlo simulation which models the interaction of photons and cosmic rays with the atmosphere and the response of the detector to the extensive air showers they generate. The effective area $A(E,\theta)$ is equated to the ratio of the number of events that satisfies a given set of cuts to the total number of events multiplied by the total throw area of the Monte Carlo simulation. The cuts are comprised of a trigger cut, an angle cut and a gamma-hadron separation cut. For the trigger cut, HAWC accepts events with the number of PMTs hit by the air shower, $nHit$, greater than a certain value. The angle cut is used to specify the direction of the photons and is a measure of HAWC's angular resolution. In the simulated events, an angular parameter defined to be the difference between the true location of the particle in the sky and the reconstructed sky location of the particle, $DelAngle$, is used as a proxy for the angular search bin-size.

Because we seek the sensitivity in the case where there is no prior knowledge of the burst location, we need to take into account the number of trials performed for the search. The optimum spatial bin-size depends on the search duration, the trigger criteria, and the value of the gamma-hadron separation parameter. The number of time bins is estimated by dividing the total search period (estimated as 5 years for HAWC) by the burst duration $\tau$. Thus the total number of trials depends on the burst duration $\tau$, the optimal spatial bin-size $DelAngle$, the trigger criterion $nHit$ and the value of the compactness parameter. In order to find the optimum set of cuts, we performed a simple parameter search and identified the set of values which give the strongest PBH burst rate upper limit. We refer the reader to reference~\cite{Ukwatta2015} for more details on the optimization process and the background rate calculations.

Figure~\ref{pbh_limits} displays the PBH Burst Rate Density Upper Limits projected for HAWC, compared
with limits from previous direct search experiments. The HAWC projected PBH rate density limit is
strictest around 10 seconds. The general features of Figure~\ref{pbh_limits},
which shows limits as a function of $\tau$, can be understood as follows.
From Equations \ref{distanceEq}--\ref{ul99}, we can see that $UL$, the upper
limit on the local PBH rate density, scales as

\begin{equation}\label{ul}
UL \sim (\mu_0(\tau)/N_{\gamma}(\tau)) ^{3/2}
\end{equation}
\\
where $\mu_0$ is the sensitivity for a given search bin $\tau$, and $N_{\gamma}(\tau)$
is the number of observable photons produced by the source.  Better sensitivity corresponds
to smaller $\mu_0$, the number of signal photons required for detection of a signal, and a stricter
PBH limit.  For a source at a given distance (that is, at the outer edge of the
volume considered), $N_{\gamma}(\tau)$ decreases when the search interval is smaller, and
produces a weaker PBH limit. Shorter intervals incur less background events but fewer source
photons. In this case, $\mu_0$ is dominated by statistical fluctuations, in
particular those associated with the detector background rate, while $N_{\gamma}(\tau)$
is dominated by the PBH emission time profile, slightly modified by the energy-dependent effective
area of the detector. These dependencies are both power laws, as seen in Figure 1 of
reference~\cite{Ukwatta2015b}, and, despite their very different physical origins, nearly cancel.
Secondary effects such as the larger number of trials incurred for shorter intervals (proportional to $1/\tau$),
and the ability to optimize background rates for larger $\tau$ for which the detection
efficiency is higher, give the residual $\tau$ dependence seen in our Figure~\ref{pbh_limits}.

According to our study, if a single PBH explodes within 0.074 parsecs of Earth and within
26 degrees of zenith, HAWC will have a 95\% probability of detecting it at 5$\sigma$ (as optimized
in a 10 s search after trials corrections). Also, HAWC would see with 95\% probability a PBH
burst within 37 degrees of zenith if it happens within 0.058 parsec of Earth~\cite{Ukwatta2015}.

The HAWC estimated sensitivity presented above was based on choosing a single set of selection criteria
(and gamma-hadron selection) for each search window duration.  We have begun exploring improvements which
might be obtained by making these selection criteria depend on $nHit$, the number of PMTs participating in
the event, as is used in current HAWC analyses~\cite{Pretz2015}. We evaluated HAWC's PBH sensitivity using
$nHit$-dependent cuts tuned for these other HAWC analyses and see a factor of 1.3 improvement in the expected
PBH burst-rate-density limit for 1 and 10 s search durations. Once the cut optimization is performed for a full
likelihood search incorporating knowledge of the PBH time and energy dependence discussed in reference~\cite{Ukwatta2015b},
we may obtain a larger improvement.

\section*{Acknowledgments}
\footnotesize{
We acknowledge the support from: the US National Science Foundation (NSF);
the US Department of Energy Office of High-Energy Physics;
the Laboratory Directed Research and Development (LDRD) program of
Los Alamos National Laboratory; Consejo Nacional de Ciencia y Tecnolog\'{\i}a (CONACyT),
Mexico (grants 260378, 55155, 105666, 122331, 132197, 167281, 167733);
Red de F\'{\i}sica de Altas Energ\'{\i}as, Mexico;
DGAPA-UNAM (grants IG100414-3, IN108713,  IN121309, IN115409, IN111315);
VIEP-BUAP (grant 161-EXC-2011);
the University of Wisconsin Alumni Research Foundation;
the Institute of Geophysics, Planetary Physics, and Signatures at Los Alamos National Laboratory;
the Luc Binette Foundation UNAM Postdoctoral Fellowship program.
}

\bibliography{icrc2015-0710}

\providecommand{\href}[2]{#2}\begingroup\raggedright\begin{thebibliography}{10}

\bibitem{Carr2010}
B.~J. {Carr}, K.~{Kohri}, Y.~{Sendouda}, and J.~{Yokoyama}, {\it {New
  cosmological constraints on primordial black holes}},  {\em \prd} {\bf 81}
  (May, 2010) 104019, [\href{http://arxiv.org/abs/0912.5297}{{\tt
  arXiv:0912.5297}}].

\bibitem{BH_Hawking}
S.~W. {Hawking}, {\it {Black hole explosions?}},  {\em \nat} {\bf 248} (Mar.,
  1974) 30--31.

\bibitem{MacGibbon1990}
J.~H. {MacGibbon} and B.~R. {Webber}, {\it {Quark- and gluon-jet emission from
  primordial black holes: The instantaneous spectra}},  {\em \prd} {\bf 41}
  (May, 1990) 3052--3079.

\bibitem{MacGibbon2008}
J.~H. {MacGibbon}, B.~J. {Carr}, and D.~N. {Page}, {\it {Do evaporating black
  holes form photospheres?}},  {\em \prd} {\bf 78} (Sept., 2008) 064043,
  [\href{http://arxiv.org/abs/0709.2380}{{\tt arXiv:0709.2380}}].

\bibitem{PageHawking1976}
D.~N. {Page} and S.~W. {Hawking}, {\it {Gamma rays from primordial black
  holes}},  {\em \apj} {\bf 206} (May, 1976) 1--7.

\bibitem{Wright1996}
E.~L. {Wright}, {\it {On the Density of Primordial Black Holes in the Galactic
  Halo}},  {\em \apj} {\bf 459} (Mar., 1996) 487,
  [\href{http://arxiv.org/abs/astro-ph/9509074}{{\tt astro-ph/9509074}}].

\bibitem{Abe2012}
K.~{Abe}, H.~{Fuke}, S.~{Haino}, T.~{Hams}, M.~{Hasegawa}, A.~{Horikoshi},
  K.~C. {Kim}, A.~{Kusumoto}, M.~H. {Lee}, Y.~{Makida}, S.~{Matsuda},
  Y.~{Matsukawa}, J.~W. {Mitchell}, J.~{Nishimura}, M.~{Nozaki}, R.~{Orito},
  J.~F. {Ormes}, K.~{Sakai}, M.~{Sasaki}, E.~S. {Seo}, R.~{Shinoda}, R.~E.
  {Streitmatter}, J.~{Suzuki}, K.~{Tanaka}, N.~{Thakur}, T.~{Yamagami},
  A.~{Yamamoto}, T.~{Yoshida}, and K.~{Yoshimura}, {\it {Measurement of the
  Cosmic-Ray Antiproton Spectrum at Solar Minimum with a Long-Duration Balloon
  Flight over Antarctica}},  {\em Physical Review Letters} {\bf 108} (Feb.,
  2012) 051102, [\href{http://arxiv.org/abs/1107.6000}{{\tt arXiv:1107.6000}}].

\bibitem{Glicenstein2013}
{\bf H.E.S.S.} Collaboration, J.~{Glicenstein}, A.~{Barnacka}, M.~{Vivier}, and
  T.~{Herr}, {\it {Limits on Primordial Black Hole evaporation with the
  H.E.S.S. array of Cherenkov telescopes}},  in {\em Proc. 33th ICRC}, (Rio de
  Janeiro, Brazil), August, 2013.

\bibitem{Pretz2015}
{\bf HAWC} Collaboration, J.~Pretz, {\it {Highlights from the High Altitude
  Water Cherenkov Observatory}},  in {\em Proc. 34th ICRC}, (The Hague, The
  Netherlands), August, 2015.

\bibitem{Smith2015}
{\bf HAWC} Collaboration, A.~Smith, {\it {HAWC: Design, Operation,
  Reconstruction and Analysis}},  in {\em Proc. 34th ICRC}, (The Hague, The
  Netherlands), August, 2015.

\bibitem{MacGibbon1991}
J.~H. {MacGibbon}, {\it {Quark- and gluon-jet emission from primordial black
  holes. II. The emission over the black-hole lifetime}},  {\em \prd} {\bf 44}
  (July, 1991) 376--392.

\bibitem{Ukwatta2015b}
T.~N. {Ukwatta}, D.~{Stump}, J.~H. {MacGibbon}, J.~T. {Linnemann}, S.~S.
  {Marinelli}, T.~{Yapici}, and K.~{Tollefson}, {\it {Observational
  Characteristics of the Final Stages of Evaporating Primordial Black Holes}},
  {\em ArXiv e-prints} (July, 2015) [\href{http://arxiv.org/abs/1507.0164}{{\tt
  arXiv:1507.0164}}].

\bibitem{Petkov2008}
V.~B. {Petkov}, E.~V. {Bugaev}, P.~A. {Klimai}, M.~V. {Andreev}, V.~I.
  {Volchenko}, G.~V. {Volchenko}, A.~N. {Gaponenko}, Z.~S. {Guliev}, I.~M.
  {Dzaparova}, D.~V. {Smirnov}, A.~V. {Sergeev}, A.~B. {Chernyaev}, and A.~F.
  {Yanin}, {\it {Searching for very-high-energy gamma-ray bursts from
  evaporating primordial black holes}},  {\em Astronomy Letters} {\bf 34}
  (Aug., 2008) 509--514, [\href{http://arxiv.org/abs/0808.3093}{{\tt
  arXiv:0808.3093}}].

\bibitem{Ukwatta2015}
{\bf HAWC and Milagro} Collaboration, A.~A. {Abdo}, A.~U. {Abeysekara},
  R.~{Alfaro}, B.~T. {Allen}, C.~{Alvarez}, J.~D. {{\'A}lvarez}, and {et. al.},
  {\it {Milagro limits and HAWC sensitivity for the rate-density of evaporating
  Primordial Black Holes}},  {\em Astroparticle Physics} {\bf 64} (Apr., 2015)
  4--12, [\href{http://arxiv.org/abs/1407.1686}{{\tt arXiv:1407.1686}}].

\end{thebibliography}\endgroup

\end{document}